# Acoustic Analogues of High-Index Optical Waveguide Devices

Farzad Zangeneh-Nejad and Romain Fleury*

*Laboratory of Wave Engineering, EPFL, 1015 Lausanne, Switzerland*

*To whom correspondence should be addressed. Email: romain.fleury@epfl.ch*

**High index optical waveguide devices such as slab waveguides, strip waveguides and fibers play extremely important roles in a wide range of modern applications including telecommunications, sensing, lasing, interferometry, and resonant amplification. Yet, transposing these advantageous applications from optics to acoustics remains a fundamental practical challenge, since most materials exhibit refractive indices lower than that of air for sound waves. Here, we demonstrate the relevance of acoustic metamaterials for tackling this pivotal problem. More specifically, we consider a metamaterial built from subwavelength air-filled acoustic pipes engineered to effectively exhibit a higher refractive index than homogenous air. We show that such medium can be employed to realize acoustic equivalents of dielectric slab or strip waveguides, and optical fibers. Unlike conventional acoustic pipes, our guiding approach allows the waveguide to remain open to the external medium, which opens a plethora of new opportunities in noise management, medical imaging, underwater communication systems, and sensing.**

**Introduction**

Guiding of light by refraction was historically demonstrated by the seminal work of Jean-Daniel Colladon in the early 1840s. In his inspiring experiment, he elegantly exploited total internal reflection between water and air to guide light through a stream of water flowing in a parabolic trajectory, or what he called the "Light Fountain" [1]. His experiment indeed formed the basic principle of many modern high index optical waveguide devices such as slab waveguides [2], strip waveguides [3], optical fibers [4] and optical cavities [5]. Such components found their widespread real-word applications not only in telecommunication [6,7], but also in sensing [8,9], lasing [10,11], amplification [12], nonlinear optics [13], and novel optical information processing technologies [14].

A typical high-index optical waveguide consists of two different layers of dielectric materials, known as core and cladding. The core layer has a higher refractive index (i.e. a lower speed of light) than the cladding. The underlying waveguiding principle then relies on total internal reflection at the boundary between the core and cladding: when light traveling in the core strikes the boundary of the cladding at an angle larger than the so-called critical angle, it is totally reflected. Light is therefore trapped inside the core, and is guided by bouncing repeatedly off the cladding [15].

A unique advantage of such high-index components, besides their large bandwidth, compactness and low attenuation is that they remain open and sensitive to the external world, i.e. the interior and exterior fields are not independent, like in the case of metallic rectangular waveguides. This feature enables various interesting and advantageous possibilities, including resonant tunneling of an externally incident beam [16], direct coupling from the far-field to a guided mode by the addition of subwavelength scatterers in the near-field [17], sensitivity to the presence of adsorbed molecules [18], and external pumping to trigger non-linear phenomena [19]. Such type of "open" waveguide, however, is completely missing in airborne acoustics, as the guiding mechanism has been traditionally contigent on sound channeling through pipes surrounded with acoustically hard walls, drastically limiting the toolkit of acousticians when it comes to designing innovative broadband noise management systems like far-field noise sinks and deflectors, energy harvesters, non-linear audio processing devices, sensors, and underwater communication systems. The absence of acoustic analog for high-index guiding stems from a pivotal and seemingly insurmountable obstacle: sound guiding by total internal reflection requires having a medium (core), inside which the speed of sound is lower than that of the surrounding air. Unfortunately, in contrast to optics, there exist no natural solid material possessing this key property [20].

The recent advances in the field of metamaterials, however, have paved the way to tackle this problem. Metamaterials are artificially structures composed of subwavelength inclusions engineered to exhibit properties that are not readily found in nature [21-43]. Over the past decade, these so called man-made materials have been widely employed to demonstrate exotic acoustic functionalities like negative refraction [44-47], perfect focusing [48-52], imaging [30-55] and cloaking [56-59]. More to the point of our interest here, some generations of them have also been exploited to slow down airborne audible sound [60-63]. To date, however, these advances have been mostly limited to physics-driven explorations centered on subwavelength sound manipulation or attenuation, without successfully translating them from basic science to technology-enabling, engineering-oriented applications such as waveguiding.

Here, we demonstrate the acoustic analogs of high-index optical devices, enabling sound guiding by open metamaterial waveguides. To this end, we first design and demonstrate an acoustic metamaterial exhibiting an effective refractive index higher than air, a crucial requirement for trapping sound and guiding it by total internal reflection. Taking advantage of the latter, we then prove how this metamaterial enables the realization of various acoustic waveguide devices including slab waveguides, strip waveguides, acoustic fibers and cavities.

Our proposal is to our knowledge unique in leveraging acoustic metamaterials to demonstrate such promising devices.

**Results**

Consider first a two-dimensional crystal with square lattice, built from air-filled acoustic pipes (Fig. 1a). Shown in Fig. 1b is a single unit cell of the crystal. The lattice constant is *a=5 cm*, and the width of the channels is chosen to be *d=0.3a*. Similar to labyrinthine metamaterials [60-63], such a lattice structure provides larger propagation paths for sound than in homogenous air, effectively reducing its travelling velocity. The calculated band structure of the crystal, shown in Fig. 1c, constitutes an evidence of this salient feature, where the lowest band is indeed seen to fall below the sound cone. The factor by which the sound velocity is decreased with respect to $c_0$ (sound speed in air), can be pictured as an effective refractive index. In fact, as illustrated in the inset of Fig. 1c, the dispersion of the lowest band is linear over a broad bandwidth ranging from 0 to 1.5 kHz (approximated by the red dashed line). The slope of this linear fit is nothing but $c_0/(2\pi n_{eff})$, where $n_{eff}$ is the refractive index the metamaterial effectively exhibits. Remarkably, this effective refractive index is higher than that of homogenous air, ($n_{eff} = 1.27 > 1$). This constitutes a key step towards trapping the sound inside the metamaterial and guiding it. We further note that the observed linear behavior of frequency dispersion is isotropic, consistent with the previous observations [64]. In fact, the three-dimensional dispersion of the first band shown in Fig. 1d is close to a cone in the wave-vector space, and the band lies below the sound cone in all directions.

To prove how the metamaterial under study enables synthesis of high index acoustic waveguides, we first investigate the realization of a slab waveguide, the simplest yet most illustrative type of such devices. Consider the configuration of Fig. 2a, where a finite piece of our metamaterial (*L=11a*) is placed in air. Despite the fact that the air inside the channel and the one outside are connected at the two interfaces, this high-index slab is actually capable of guiding sound via total internal reflection, similar to its optical counterpart. The condition upon which the latter happens can be obtained by an approach similar to the techniques proposed in geometrical optics [65,66]. Assume that a monochromatic sound ray is being totally reflected at the boundary between the metamaterial and air, and is travelling in a zig-zag path indicated in Fig. 2a (this is possible whenever $\theta > \theta_c = 51.8°$). To have a guided mode, the round-trip phase acquired by the ray must be an integer multiple of $2\pi$ [15], i.e.

$$\psi = 2k_0 n_{eff} L \cos\theta - 4\phi_r = 2\upsilon\pi \tag{1}$$

where $k_0$ is the free space wave number, $2\phi_r$ is the phase shift on total internal reflection and $\upsilon$ is an integer. Considering the fact that the ray is effectively travelling along $y$ with the propagation constant of $\beta = k_0 n_{eff} \sin(\theta)$, condition (1) actually determines the dispersion relation of guided modes, i.e. the allowed values for $\beta$ at a given frequency. A more rigorous approach based on wave analysis can also be employed to obtain the dispersion relation of Eq. 1 (see Methods).

For instance, let us assume the frequency of the trapped ray to be $f_0 = 1\ kHz$. Fig. 2b reports the variation of $\psi$ versus $\theta$. Since $\psi/2\pi$ must be an integer, we can adequately determine the set of incident angles $\theta_i$, or equivalently propagation constants $\beta_i$, for which we have guided modes. As observed, at the incident angles of $\theta_1 \approx 80°$, $\theta_2 \approx 68°$ and $\theta_3 \approx 56°$, corresponding to the propagation constants of $\beta_1 = 1.2507 k_0$, $\beta_2 = 1.1775 k_0$ and $\beta_3 = 1.0566 k_0$, $\psi$ has become an integer multiple of $2\pi$ satisfying the dispersion relation of Eq. 1. These values of $\beta_1$, $\beta_2$ and $\beta_3$ are indeed the propagation constants of the first, second and third order modes of our acoustic slab waveguide, whose profiles are found independently from finite-elements simulations and represented in Figs. 2c, d and e, respectively. Inspection of these figures reveals how sound is confined within the metamaterial, despite the fact that the slab is acoustically open at its interfaces. As a matter of fact, in perfect analogy with a typical dielectric slab waveguide, the successive mode profiles involve even and odd sinusoidal distributions in the middle layer, whereas they exponentially decay in the two surrounding regions. The results of these three figures illustrate how the guiding properties are dictated by the metamaterial structure rather than from its components. This effective medium description vindicates our choice for the sound trajectory in Fig. 2a, which does not follow the directions of the pipes.

The proposed acoustic slab waveguide guides sound properly, but provides confinement only in one direction. We now further investigate the realization of two dimensional waveguides providing confinement in both transverse directions. Consider the geometry of Fig. 3a, which is an acoustic strip waveguide constructed by truncating the metamaterial to a rectangular cross section. Notice that the way the waveguide is truncated affects the macroscopic wave modes but not the high-index property of the metamaterial. Unlike slab waveguides, these waveguides do not possess an analytical solution for guided modes. Nonetheless, there are approximate methods [3,67], by which their frequency dispersion can be properly estimated. Here, we take an approach proposed by Marcatili [3], while the Methods section reports a more advanced model. Assume the structure is infinite along $z$ direction. The wave equation for a guided mode traveling with the propagation constant of $\beta$ can then be expressed by

$$\frac{\partial^2 P}{\partial x^2} + \frac{\partial^2 P}{\partial y^2} + ((k_0 n(x,y))^2 - \beta^2)P = 0 \qquad (2)$$

Where $P$ is pressure, $k_0$ is the free space wavenumber, and $n(x, y)$ is the refractive index profile. The key underlying principle of Marcatili's method is that it neglects the pressure fields in the regions I to IV (see Fig. 3a), an assumption which is not very restrictive as the field exponentially decays in these regions. Using this trick and considering the symmetry of the structure with respect to *x* and *y* axes, one may envision the following form of solution for the pressure fields in other regions

$$P = \begin{cases} P_0 \cos(k_{f,x} x - \varphi_x) \cos(k_{f,y} y - \varphi_y) & 0 < x < L_x, 0 < y < L_y \\ P_0 \cos(k_{f,x} L_x - \varphi_x) \cos(k_{f,y} y - \varphi_y) e^{-\gamma_{c,x}(x-L_x)} & x > L_x, 0 < y < L_y \\ P_0 \cos(k_{f,x} x - \varphi_x) \cos(k_{f,y} L_y - \varphi_y) e^{-\gamma_{c,y}(y-L_y)} & 0 < x < L_x, y > L_y \end{cases} \quad (3)$$

where the wave numbers $k_{f,x}$, $k_{f,y}$, $\gamma_{c,x}$, and $\gamma_{c,y}$ satisfy

$$\begin{cases} k_{f,x}^2 + k_{f,y}^2 = k_0^2 n_{eff}^2 - \beta^2 \\ \gamma_{c,x}^2 = \beta^2 + k_{f,y}^2 - k_0^2 \\ \gamma_{c,y}^2 = \beta^2 + k_{f,x}^2 - k_0^2 \end{cases} \quad (4)$$

We note that based on the observed symmetry, one would expect the pressure distribution to be either even or odd with respect to the origin. This imposes the following restrictions on the phases $\varphi_x$ and $\varphi_y$

$$\begin{aligned} \varphi_x &= (X-1)\frac{\pi}{2} \quad X = 1, 2, \ldots \\ \varphi_y &= (Y-1)\frac{\pi}{2} \quad Y = 1, 2, \ldots \end{aligned} \quad (5)$$

Applying the continuity of particle velocity at *x=L$_x$* and *y=L$_y$* yields

$$\begin{aligned} k_{f,x} L_x &= (X-1)\frac{\pi}{2} + \tan^{-1}(\frac{n_{eff}^2 \gamma_{c,x}}{k_{f,x}}) \quad X = 1, 2, \ldots \\ k_{f,y} L_y &= (Y-1)\frac{\pi}{2} + \tan^{-1}(\frac{n_{eff}^2 \gamma_{c,y}}{k_{f,y}}) \quad Y = 1, 2, \ldots \end{aligned} \quad (6)$$

By solving Eqs. 4 and 6 simultaneously, one can conveniently determine the set of propagation constants $\beta$ related to guided modes. Here, we assume $2L_x = 2L_y = 11a$, and focus first on the lowest order mode (setting $X = Y = 1$) at the frequency of $f_0 = 1$ kHz, obtaining $\beta = 1.2122 k_0$. Fig. 3b (top panel) shows the field profile of this mode obtained independently from numerical calculations, from which one can observe how sound is confined in both transverse directions. Notably, the field pattern observed in this figure is not only consistent with our expectation based on the pressure distribution in Eq. 3, but the propagation constant obtained via full-wave simulations is consistent with the one obtained from Marcatili's method. The field pattern of the corresponding second order mode is also provided in Fig. 3b (bottom panel). Note that these two modes are degenerate, i.e. both correspond to the same propagation constant. The two-dimensional confinement achieved here is highly promising for various applications such as interferometry [68] and amplification [65].

As another important case, we demonstrate the acoustic counterpart of an optical fiber. To this end, consider again a finite piece of the metamaterial, now truncated to a circular cross section as depicted in Fig. 3c. An analysis similar to that previously considered in the case of the rectangular cross-section yields the following dispersion relation:

$$k_f \frac{J_m'(k_f R)}{J_m(k_f R)} = \gamma_c n_{eff}^2 \frac{K_m'(\gamma_c R)}{K_m(\gamma_c R)} \tag{7}$$

in which $k_f = \sqrt{k_0^2 n_{eff}^2 - \beta^2}$, $\gamma_c = \sqrt{\beta^2 - k_0^2}$, $J_m$ is mth order Bessel function of first kind and $K_m$ is that of second kind. From Eq. 7, we calculate the propagation constants of the first and second order modes to be $\beta_1 = 1.2157 k_0$ and $\beta_2 = 1.1158 k_0$, respectively (for $R = 15a$ and $f_0 = 1 KHz$). The profiles corresponding to these modes are also numerically calculated and plotted in the insets of Fig. 3d. Remarkably, these field patterns are akin to LP$_{01}$ and LP$_{11}$ mode profiles in optical fibers (see Supplementary Fig. 1 for third order mode, i.e. LP$_{21}$). We note that the two LP$_{11}$ modes are actually degenerate and have the same propagation constants. It is also remarkable that albeit such acoustic fibers remain open to the outside air at their boundaries, the modes are guided without leaking in the external medium.

We finally move onto the case of designing an open acoustic cavity, capable of confining sound in all three dimensions. Consider the three-dimensional generalization of our two dimensional metamaterial (Fig. 4a), i.e. a cubic array of acoustic pipes filled with air (the medium between the pipes is assumed to be a much denser solid material, not represented in Fig. 4a). Like the previous cases, the obtained band structure (Fig. 4b) illustrates how the structure is imitating a medium with an effective refractive index higher than air, over a broad frequency range from 0 to 1.5 kHz. As a result, when surrounded by air for example in Fig. 4c, this cube behaves like an acoustic cavity, which supports acoustic resonances of finite life-times (the absence of total internal reflection in this case makes the mode leaky). This anticipation can be conveniently validated from the result of our full-wave numerical simulation in Fig. 4d illustrating the lowest order resonance, which occurs at 944.3 Hz and is associated with a radiative decay rate of 38.27 Hz.

**Experimental validation**

Based on these theoretical investigations we built a prototype of a finite size high-index metamaterial, whose photograph is shown in Fig. 5a. The device consists of a sample of rubber-like polymer, perforated by an array of holes taking the role of the acoustic pipes (Fig. 5b). The length, width and thickness of the fabricated device are $L_1=47.25$ cm, $L_2=27.5$ cm, and $a=2.25$ cm, respectively. Moreover, the radii of the air channels are assumed to be $r=0.5$ mm. Since the wavelength of operation is around $\lambda_0 \approx 30 cm$, the structuration of the sample is deeply subwavelength. To evidence the resonances of the high index metamaterial, we excite the sample from its back side with a loudspeaker, and measure the pressure field transmitted to the other side with a microphone as shown in Fig. 5c (see Methods for more details regarding the setup employed here). The measured transmission spectrum is then reported in Fig. 5d. As observed, the spectrum has a peak around frequency $f_g=1.45$ kHz. This corresponds to resonant tunneling through the first cavity mode, as numerically evidenced by direct finite-element simulations, which predict a first dipolar resonance frequency at $f=1.4$ kHz (Fig. 5e). We eventually investigate the profile of this mode in experiment as well, by imaging the pressure field in the near field of the sample (see Methods). The measured field profile depicted in Fig. 5f is in perfect agreement with our numerical simulations of Fig. 5e, demonstrating that the fabricated device indeed behaves like a high-index open acoustic cavity.

We further provide a very simple and intuitive experiment in order to elaborate how the idea presented in the manuscript enables open-type waveguiding. Consider the setup shown in Fig. 6a, consisting of four distinct acoustic pipes with the diameters of *2 cm*, made of glass and arranged in a square lattice. As described previously, these obstructing tubes of subwavelength cross-section create an acoustic equivalent of high-index optical fibers. The sound generated by the loudspeaker is therefore effectively slowed down as it travels in this waveguide (in the air between the tubes). More importantly, the energy remains confined and avoids radiating to other directions. To demonstrate these expectations, we have used three microphones, labeled by different numbers in Fig. 6a, so as to measure the pressure field at different locations. Fig. 6b represents the pressure measured by the microphone 1, 2 and 3. Comparing the pressure fields measured by the first and second microphones reveals that the sound is properly guided between the tubes. Notably, the time delay between the two pulses is 0.78 *msec* (measured between the two signal precursors), which is higher than the time it takes for the sound to travel between these two microphones in free-air (0.62 *msec*). This clearly demonstrates a reduction of the sound speed by 14.7 percent in the slow-wave property of the configuration under study. Moreover, the low pressure level measured by the third microphone (which is placed outside the tubes) evinces the low radiation of the sound to unwanted directions. Shown in Fig. 6c are the pressure fields measured by each of the microphones when the tubes are removed but the microphones and source are kept at the same places. It is clear that the amounts of pressure measured by the microphones are almost equal. Therefore, there is no field confinement in this case even though the sound is being propagated in free-air.

**Discussion and Conclusion**

To conclude, we have demonstrated the acoustic equivalents of high index optical waveguide devices, including slab waveguides, strip waveguides, acoustic fibers and cavities. This is achieved in simple holey acoustic metamaterials engineered to effectively exhibit a higher refractive index than homogenous air. Our numerical simulation and experimental measurements highlighted the analogies between such devices and their optical counterparts. We envision the generation of many novel devices utilizing the sound confinement provided by such high index acoustic components. As a tangible example, we point out how such high-index devices we presented can be leveraged for realizing a simple acoustic sensor. Consider again the experimental setup of Fig. 5c. Any change in the ambient air properties, will vary the resonance frequency of the cavity mode. This in turn results in a shift in the transmission spectrum of the sample, changing the frequency at which the transmission reaches its maximum value in Fig. 5d. Using this effect, the ambient basic properties such as its humidity or temperature can be continuously sensed and monitored via our acoustic setup. The sensitivity of the sensor can be optimized by employing more advanced designs involving Fano resonances, for instance. The fact that such waveguides and cavities are open to the external medium is a key feature of our findings: coupling with external flow, for instance, combined with slower wave materials such as with labyrinthine paths, may allow the next generation of two-port non-reciprocal acoustic devices [69,70]. Placing subwavelength scatterers in the near field of the slab can allow broadband coupling of the guided mode to the far field, an interesting possibility for noise control and management. Other promising and appealing devices for signal processing applications, including sound amplifiers, non-linear switches, acoustic gratings, isolators, filters and modulators can also be envisioned to be realized utilizing these waveguides. Altogether, we believe that our proposal can dramatically expand the toolkit of acoustic engineering.

**Methods**

**Wave analysis for the acoustic slab waveguide**

We provide here wave analysis for the proposed acoustic slab waveguide, whose dispersion relation was already obtained in Eq. 1 using a simple geometrical approach. Consider again the configuration of Fig. 2a. Assuming that the structure is infinite along the $y$ and $z$ direction, we can write the wave equation as

$$\frac{\partial^2 P}{\partial x^2} + ((k_0 n(x))^2 - \beta^2)P = 0 \tag{8}$$

where $P$ is the pressure field travelling along $y$ with the propagation constant of $\beta$, $k_0$ is the free space wavenumber and $n(x)$ is the refractive index. We are interested in guided modes, which are confined in the middle region and exponentially decay outside. We therefore envision the following solution for the above wave equation

$$P = \begin{cases} P_1 e^{-\gamma_c (x-L)} & x > L \\ P_0 \cos(k_f x - \phi_r) & 0 < x < L \\ P_1 e^{\gamma_c x} & x < 0 \end{cases} \tag{9}$$

in which $\gamma_c = \sqrt{\beta^2 - k_0^2}$ and $k_f = \sqrt{k_0^2 n_{eff}^2 - \beta^2}$. The latter pressure field distribution should satisfy the corresponding boundary conditions, which are the continuity of pressure and velocity at each interface, that is $x=0$ and $x=L$. Matching pressure at $x=0$ leads to

$$P_1 = P_0 \cos(\varphi_r) \tag{10}$$

while matching the normal particle velocity yields

$$n_{eff}^2 \gamma_c P_1 = k_f P_0 \sin(\varphi_r) \tag{11}$$

We notice that the continuity of the y components of particle velocities results in the same relation as Eq. 10. Dividing Eq. 11 to Eq. 10, we can get rid of field amplitudes and obtain an expression for $\varphi_r$ as follows

$$\tan(\varphi_r) = \frac{n_{eff}^2 \gamma_c}{k_f} \tag{12}$$

Similarly, the continuity of pressure and velocity at $x=L$ lead to the following equation

$$\tan(k_f L - \varphi_r) = \frac{n_{eff}^2 \gamma_c}{k_f} \tag{13}$$

Taking inverse tangent of the latter results in

$$k_f L - \varphi_r + \upsilon\pi = \tan^{-1}(\frac{n_{eff}^2 \gamma_c}{k_f}) \tag{14}$$

But according to Eq. 12, $\tan^{-1}(\frac{n_{eff}^2 \gamma_c}{k_f}) = \varphi_r$. This then turns out the following relation

$$k_f L - 2\varphi_r = \nu\pi \tag{15}$$

Note that $2\varphi_r$ in the above equations is nothing but the phase on total internal reflection acquired by the sound ray in Fig. 2a. Considering the fact the ray is effectively travelling along y direction with the propagation of $\beta = k_0 n_{eff} \sin(\theta)$, one can easily deduce $k_f = k_0 n_{eff} \cos(\theta)$. This then reveals that the dispersion relation obtained here is exactly the same as that of Eq. 1.

**Kumar's method for estimating dispersion relation of the acoustic rectangular waveguide**

This section describes Kumar's method [67], to analytically estimate the dispersion relation of the rectangular waveguide shown again in Supplementary Fig. 2a. Similar to the previous cases, we assume the structure is infinite along $z$ direction, and start by considering the wave equation for a guided mode traveling along $z$, which is reiterated here

$$\frac{\partial^2 P}{\partial x^2} + \frac{\partial^2 P}{\partial y^2} + ((k_0 n(x,y))^2 - \beta^2)P = 0 \tag{16}$$

The method Kumar took to solve the equation relies on expressing the refractive index profile as the summation of two independent refractive indices as

$$n^2(x,y) \approx n_x^2(x) + n_y^2(y) \tag{17}$$

where $n_x^2(x)$ and $n_y^2(y)$ are expressed as

$$n_x^2(x) = \begin{cases} \dfrac{n_{eff}^2}{2} & |x| \leq L_x \\ 1 - \dfrac{n_{eff}^2}{2} & |x| > L_x \end{cases}$$

$$n_y^2(y) = \begin{cases} \dfrac{n_{eff}^2}{2} & |y| \leq L_y \\ 1 - \dfrac{n_{eff}^2}{2} & |y| > L_y \end{cases} \tag{18}$$

Notice that in the above equation, $n_x(x)$ is solely dependent on $x$ whereas $n_y(y)$ is solely dependent on $y$. This actually forms the basis to analyze the rectangular waveguide by separating it into two independent slab waveguides as shown in Supplementary Fig. 2b. To elaborate this, let us first write the pressure field as

$$P(x,y) = X(x)Y(y) \tag{19}$$

Substituting Eqs. 17 and 19 into Eq. 16 yields

$$\frac{1}{X}\frac{d^2 X}{dx^2} + \frac{1}{Y}\frac{d^2 Y}{dy^2} + ((k_0 n_x(x))^2 + (k_0 n_y(y))^2 - (\beta_x^2 + \beta_y^2)) = 0 \tag{20}$$

from which one readily calculate the propagation constant of the waveguide as $\beta^2 = \beta_x^2 + \beta_y^2$. The obtained equation can then be deliberately reduced into two independent equations as

$$\frac{d^2X}{dx^2} + ((k_0 n_x(x))^2 - \beta_x^2) X = 0$$
$$\frac{d^2Y}{dy^2} + ((k_0 n_y(y))^2 - \beta_y^2) Y = 0 \tag{21}$$

which are nothing but the wave equations for the two slabs illustrated in Supplementary Fig. 2b. Therefore, all we need to obtain the propagation constant $\beta$, is to calculate the propagation constants $\beta_x$ and $\beta_y$, corresponding to the slab waveguides of Supplementary Fig. 2b and use the relation $\beta^2 = \beta_x^2 + \beta_y^2$. It is worthy to mention that the summation of Eq. 17 results in correct values for the refractive index profile $n^2(x, y)$ in all areas except regions I to IV. In these regions the mentioned summation turns out $n(x, y) = \sqrt{2 - n_{eff}^2}$ which does not equal 1. However, it does not affect our analysis as the pressure fields are negligible in these regions.

**Wave analysis for the proposed acoustic fiber**

Here, we derive the dispersion equation for the acoustic fiber shown in Fig. 3c. As usual, we start by assuming the fiber is infinite along $z$ direction. We then write the wave equation, this time in the cylindrical coordinates as

$$\frac{\partial^2 P}{\partial r^2} + \frac{1}{r}\frac{\partial^2 P}{\partial \varphi^2} + ((k_0 n(r))^2 - \beta^2) P = 0 \tag{22}$$

Where $\beta$ is the propagation constant of a guided mode travelling along $z$, $k_0$ is the free space wave number and $n(r)$ is the refractive index profile. By solving Eq. 22 making use of separation of variables, one gets a pressure field distribution of the form

$$P = \begin{cases} P_0 \dfrac{J_m(k_f r)}{J_m(k_f R)} e^{jm\varphi} & r < R \\ P_0 \dfrac{K_m(\gamma_c r)}{K_m(\gamma_c R)} e^{jm\varphi} & r > R \end{cases} \tag{23}$$

in which $k_f = \sqrt{k_0^2 n_{eff}^2 - \beta^2}$, $\gamma_c = \sqrt{\beta^2 - k_0^2}$, $J_m$ is mth order Bessel function of first kind and $K_m$ is that of second kind. Notice that the field distribution of Eq. 23 satisfies the continuity of pressure at *r=R*. Therefore, all we need is to consider velocity continuity, leading to the following equality

$$k_f \frac{J_m'(k_f R)}{J_m(k_f R)} = \gamma_c n_{eff}^2 \frac{k_m'(\gamma_c R)}{k_m(\gamma_c R)} \qquad (24)$$

which is nothing but the dispersion relation we mentioned in Eq. 7. We notice that for the lowest order mode (*m=0*), the term $e^{jm\varphi}$ in Eq. 23 disappears, leading to a pressure field distribution independent of azimuthal direction. However, for the higher order modes, this exponential term results in two independent azimuthal components, which are $\cos(m\varphi)$ and $\sin(m\varphi)$. These two components imply two different field distributions for pressure, which are degenerate in a sense that they both correspond to the same propagation constant $\beta$. This statement is indeed consistent with the pressure field distributions we obtained in Fig. 3d and Supplementary Fig. 1.

**Full-wave simulations**

The numerical simulations throughout the whole manuscript carried out making use of COMSOL Multiphysics, acoustic module. Sound hard wall boundary conditions were applied to the edges of the acoustic pipes. The white regions in Fig. 1a were excluded from the modeling domain. In the band structure calculations, we first considered the unit cell of the metamaterial. Dispersion curves were then calculated for the infinite crystal by varying the Bloch wave numbers and performing finite-element eigenfrequency solver simulations.

**Experimental measurements**

The acoustic cavity was fabricated by Multi-Jet three-dimensional (3D) printing in a sample of TangoBlack polymer employing an Objet Connex 500 printer. The fabricated sample was then put in the acoustic setup shown in Fig. 5c. The setup includes a loudspeaker as the source, an ICP microphone as the receiver and a Data Physics Quattro acquisition system connected to a computer controlling it. The analyzer was set to make the loudspeaker generate burst noise sound between 0 and 1.5 kHz, exciting the sample from its back. An acoustic tube is also attached to the end of the loudspeaker used so as to efficiently excite the sample (see Fig. 5c). The corresponding transmitted pressure was then measured by the microphone and sent back to the analyzer. By sending the excitation signal to another channel, measuring it and setting it as the reference, the analyzer could then determine the magnitude and phase of the transmission spectrum through the sample.

In order to obtain the field profile of the lowest order resonance of the device, a Polytech PSV-500 scanning vibrometer was employed, as displayed in Supplementary Fig. 3. The vibrometer was focused on an ultrathin transparent plastic sheet placed in front of the sample. Like the previous setup, the sample was excited from its back side by pseudo-random noise. The resulting variation of acoustic pressure induced small mechanical fluctuations in the plastic plate. Such fluctuations were detected by the vibrometer, and Fourier transformed to determine the sound field amplitude at each point, creating the image of Fig. 5f.

In the second experiment, we used four acoustic pipes having the radius of *1 cm*, and made of glass. We then put them in a subwavelength lattice structure (the distance between the tubes is 6 cm which is much less than the operational wavelength $\lambda \approx 30\ cm$) so as to create an effective high-index medium. To excite the sample, we connected the loudspeaker to one of the output channels of the scanning vibrometer generating a burst type waveform at the frequency of desire. This excitation signal was also sent to another channel of the vibrometer specifying the reference signal. We then put three microphones at different locations observed in Fig. 6a and measured the resulting signals. The voltage data on the vibrometer channels were then readily transferred to the pressure levels making use of the corresponding conversion constants.

# Figures

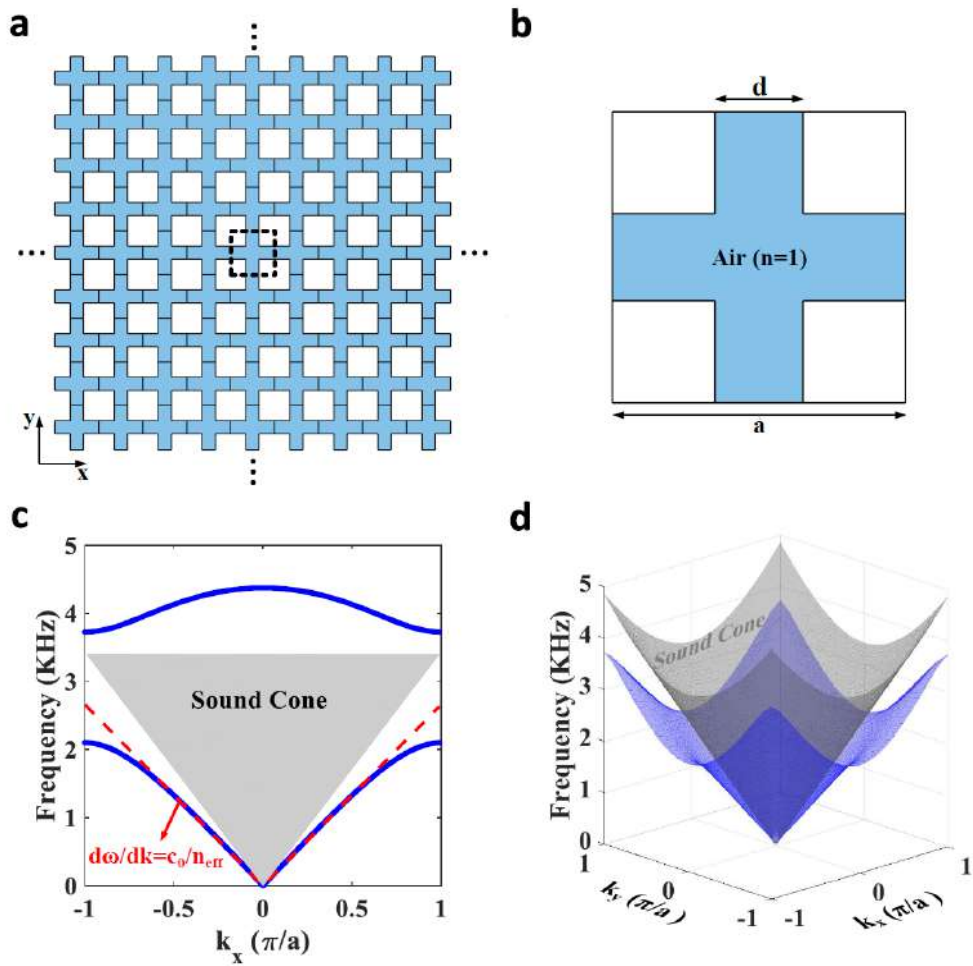

**Fig. 1: Design of an acoustic metamaterial exhibiting a refractive index higher than homogenous air. a,** Geometry of a square periodic lattice consisting of air-filled acoustic pipes. **b,** A single unit cell of the lattice shown in panel a. The unit cell length and the thicknesses of the channels are chosen to be *a=5cm* and *d=1.5 cm*, respectively. **c,** Frequency dispersion of the crystal numerically calculated by FEM. The lowest band (blue) falls below the sound cone (gray), corresponding to sound waves with velocities lower than speed of sound in air. **d,** Three-dimensional dispersion surface for the lowest band (blue) and limit of the sound cone (grey).

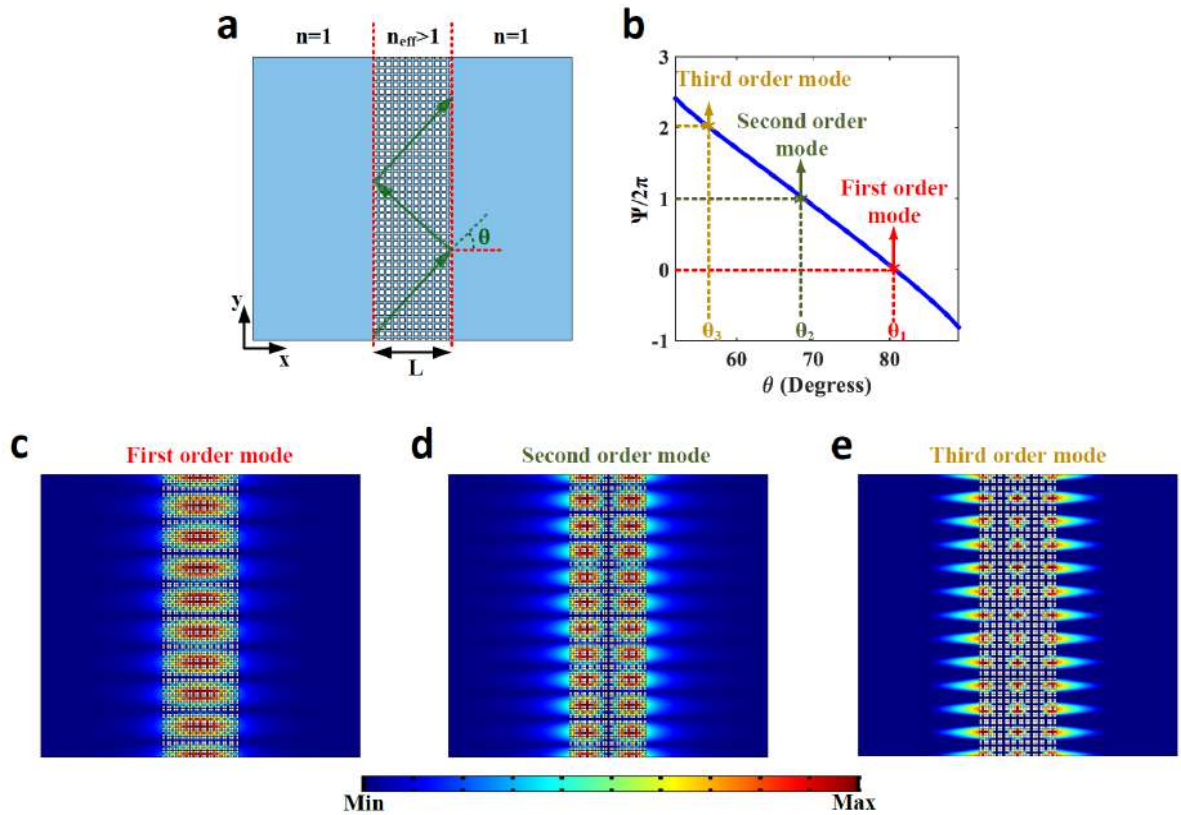

**Fig. 2: Realization of the acoustic equivalent of an optical slab waveguide by means of the metamaterial under study**, **a,** A finite piece of our metamaterial ($L=11a$) surrounded by air. A sound ray is expected to be guided by total internal reflection in a zig-zag path indicated in the figure. **b,** Variation of the round-trip phase acquired by the ray versus the incident angle θ. At the incident angles $\theta_1 \approx 80°$, $\theta_2 \approx 68°$ and $\theta_3 \approx 56°$ the round trip phase has become a multiple integer of $2\pi$, leading to guided modes. **c,** Mode profile of the first-order guided mode obtained by numerical simulations. Similar to the case of optical slab waveguides, the mode profile is sinusoidal inside the middle region whereas it exponentially decays in its surrounding. **d, e,** Same as panel c but for the second and third order modes, respectively.

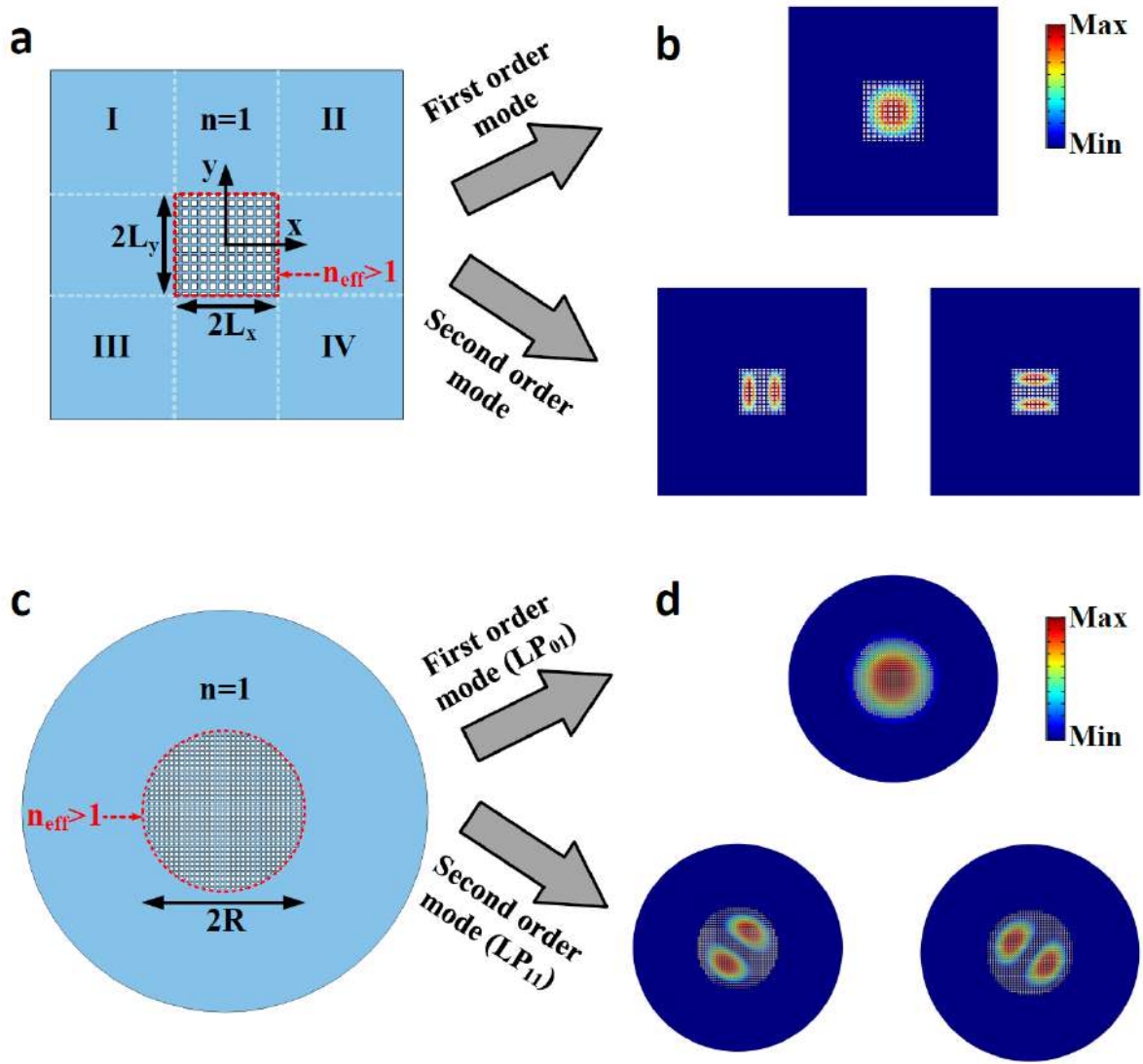

**Fig. 3: Acoustic fibers. a,** A finite piece of the metamaterial is truncated to a rectangular cross section and is surrounded by a lower index material, that is air. **b,** Mode profiles of the first (top panel) and second (bottom panel) order guided modes in the strip waveguide for *2Lx=2Ly=11a*. The guided modes have field distributions akin to optical rectangular waveguides. **c**, A finite piece of the metamaterial is truncated, this time, to a circular cross section to act as the acoustic analog of an optical fiber **d**, Profiles of the first (top panel) and second (bottom panel) order guided modes for *R=15a*. The field patterns are similar to $LP_{01}$ and $LP_{11}$ mode profiles in a real optical fiber.

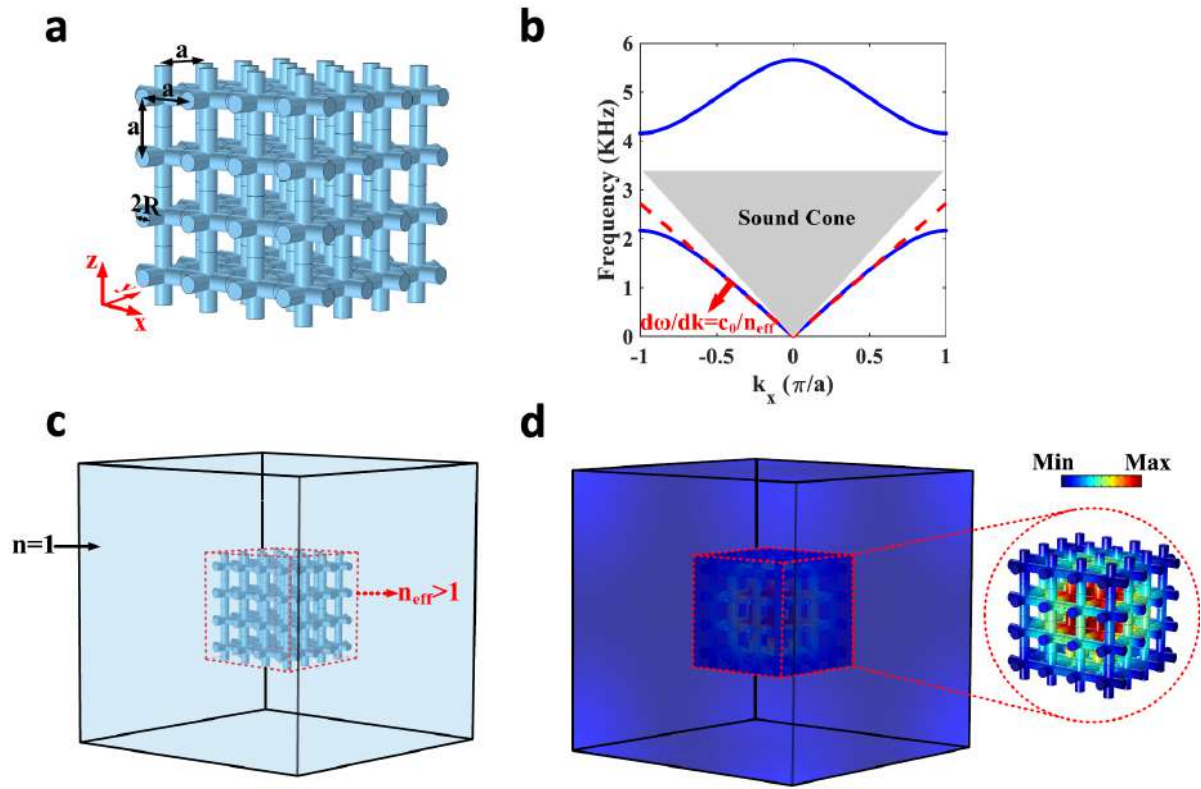

**Fig. 4: High index open acoustic cavities. a,** Three dimensional generalization of the two dimensional metamaterial: Air filled acoustic tubes with radii $2R=0.3a$ are arranged in a crystal, periodic in all three directions. **b,** Frequency dispersion of the crystal for $a=5cm$. Like the two-dimensional case, the lowest band falls outside the sound cone and can be well estimated via a linear function with a slope implying an effective refractive index higher than air for the medium **c,** A finite piece of the metamaterial is placed in air. **d,** Profile of the high-index cavity mode with finite lifetime.

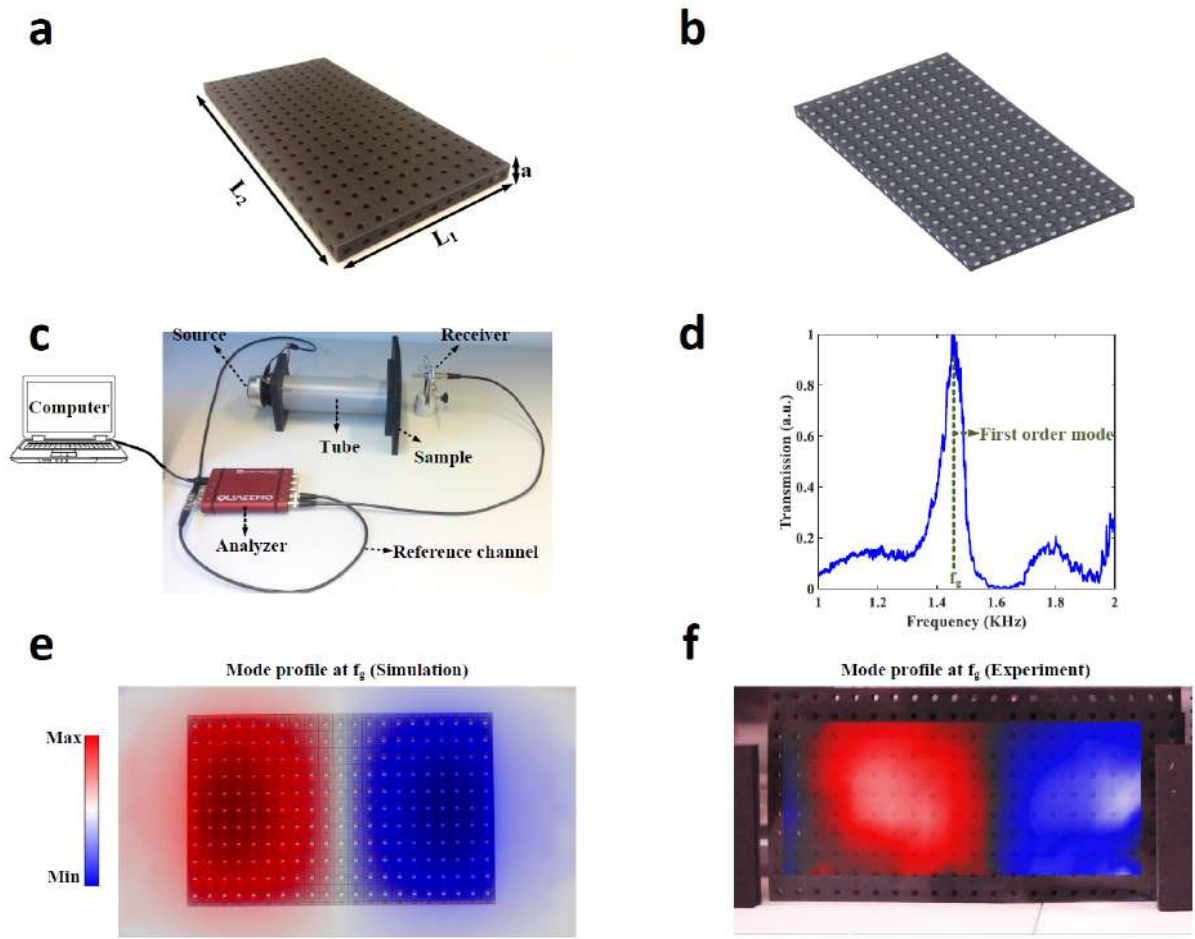

**Fig. 5: Experimental validation of the cavity mode of a high index acoustic slab. a,** An acoustic high-index metamaterial has been fabricated from rubber-like material (TangoBlack). The perforated holes inside the sample form an acoustic pipe network. **b,** Representation of the pipe network: the holes form an acoustic crystal similar to that of Fig. 2a. **c,** Acoustic setup used to evaluate the existence of acoustic guided modes supported by the sample. **d,** Pressure transmission spectrum of the fabricated sample: the spectrum is showing a peak at $f_g$=1.45 kHz, which corresponds to resonant tunneling through a cavity mode. **e,** Mode profile obtained from finite-element simulations, confirming the existence of a dipolar mode at *1.4 kHz*. **f,** Experimentally measured pressure profile above the sample, confirming the resonant excitation of the dipolar mode and the behavior of the metamaterial as a high index slab.

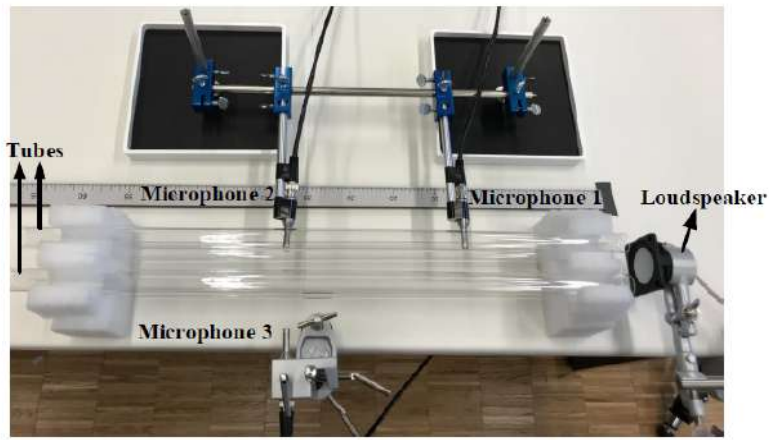

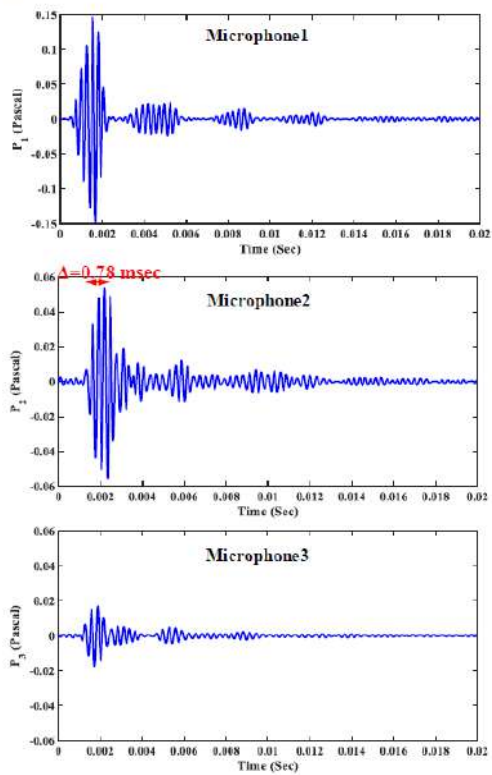
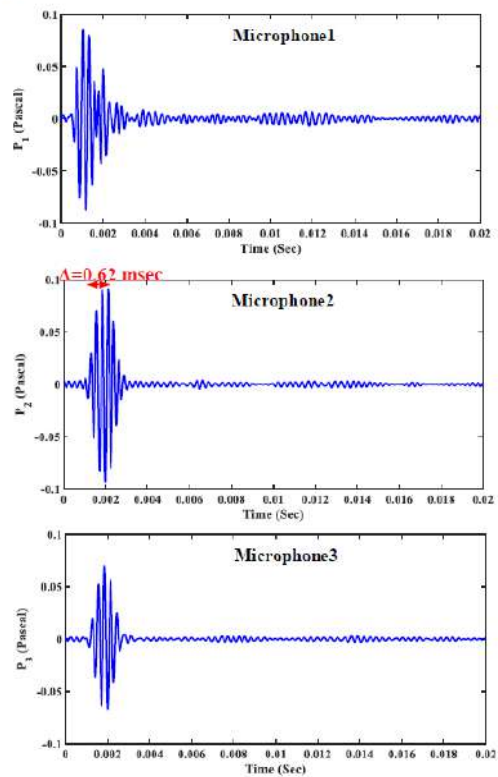

**Fig. 6: Experimental demonstration of the proposed open-type slow waveguiding scheme**: **a,** An acoustic fiber consisting of four glass pipes arranged in a square lattice is considered. The tubes effectively reduce the guided sound velocity, consistent with Fig. 3. **b,** Pressure fields measured by each of the microphones. Not only the sound is being guided along the direction of desire, but also it avoids radiation to unwanted directions. **c,** Same as panel b except that the sample is removed.

## Acknowledgments

This work was supported by the Swiss National Science Foundation (SNSF) under Grant No. 172487.

## Author Contributions

F. Z-N. carried out the research work under supervision of R.F. Both authors participated in derivation of the results and writing the manuscript.

## Competing interest

The authors declare no competing interests.

## Data availability statement

The datasets generated during and/or analyzed during the current study are available from the corresponding author on reasonable request.